\documentclass[twocolumn,prl,superscriptaddress,showpacs,floatfix,%
preprintnumbers,nofootinbib,nobalancelastpage]{revtex4}

\usepackage{graphicx}

\begin{document}

\title{Relic density of neutrinos with primordial asymmetries}

\author{Sergio Pastor}
\affiliation{Institut de F\'\i sica
 Corpuscular (CSIC-Universitat de Val\`encia),
 Ed.\ Instituts d'Investigaci\'o, Apt.\ 22085,
 46071 Val\`encia, Spain}

\author{Teguayco Pinto}
\affiliation{Institut de F\'\i sica
 Corpuscular (CSIC-Universitat de Val\`encia),
 Ed.\ Instituts d'Investigaci\'o, Apt.\ 22085,
 46071 Val\`encia, Spain}

\author{Georg G.\ Raffelt}
\affiliation{Max-Planck-Institut f\"ur Physik
(Werner-Heisenberg-Institut), F\"ohringer Ring 6, 80805 M\"unchen,
Germany}

\date{22 August 2008, revised 2 June 2009}

\preprint{MPP-2008-105, IFIC/07-59 }

%%%%%%%%%%%%%%%%%%%%%%%%%%%%%%%%%%%%%%%%%%%%%%%%%%%%%%%%%%%%%%%%%%%%%%
\begin{abstract}
We study flavor oscillations in the early universe,
assuming primordial neutrino-antineutrino asymmetries. Including
collisions and pair processes in the kinetic equations, we not only
estimate the degree of flavor equilibration, but for the first time
also kinetic equilibration among neutrinos and with the ambient
plasma. Typically the restrictive BBN bound on the
$\nu_e\bar\nu_e$ asymmetry indeed applies to all flavors as
claimed in the previous literature, but fine-tuned initial
asymmetries always allow for a large surviving neutrino excess
radiation that may show up in precision cosmological~data.
\end{abstract}
%%%%%%%%%%%%%%%%%%%%%%%%%%%%%%%%%%%%%%%%%%%%%%%%%%%%%%%%%%%%%%%%%%%%%%

\pacs{14.60.Pq, 98.70.Vc, 26.35.+c}

\maketitle

%%%%%%%%%%%%%%%%%%%%%%%%%%%%%%%%%%%%%%%%%%%%%%%%%%%%%%%%%%%%%%%%%%%%%%
{\em Introduction.}---%
%%%%%%%%%%%%%%%%%%%%%%%%%%%%%%%%%%%%%%%%%%%%%%%%%%%%%%%%%%%%%%%%%%%%%%
The state of the hot medium in the early universe before and around
big-bang nucleosynthesis (BBN) is fixed by a small number of
cosmological parameters, most importantly the baryon abundance that
nowadays is measured most accurately with precision cosmological
data. In addition there are three unknown neutrino asymmetries, one
for each flavor, that modify the cosmic neutrino density and, for the
case of the $\nu_e\bar\nu_e$ asymmetry, influence $\beta$ reactions of
the form $e^-+p\leftrightarrow n+\nu_e$ and $e^++n\leftrightarrow
p+\bar\nu_e$, and thus the primordial $^4$He
abundance~\cite{Kang:1991xa, Sarkar:1995dd, Esposito:2000hi,
  Serpico:2005bc, Pisanti:2007hk, Simha:2008mt, Iocco:2008va}.

It is often assumed that sphalerons equilibrate the lepton and baryon
asymmetries, implying that all neutrino asymmetries are of order
$10^{-9}$ or smaller, leaving no observable trace.  On the other hand,
sphalerons have never been observed, so a cosmic neutrino asymmetry
would raise fundamental questions about both electroweak physics and
the early universe. In any event, improving observational information
on neutrino asymmetries is part of the ongoing effort to determine all
cosmological parameters with increasing accuracy.

Neutrino oscillations driven by the ``solar'' and ``atmospheric'' mass
differences $\Delta m^2_{\rm
  sol}=7.65^{+0.46}_{-0.40}\times10^{-5}~{\rm eV}^2$ and $\Delta
m^2_{\rm atm}=\pm 2.40^{+0.24}_{-0.22}\times10^{-3}~{\rm eV}^2$ and by
the large mixing angles $\sin^2\theta_{12}=0.30^{+0.05}_{-0.03}$ and
$\sin^2\theta_{23}=0.50^{+0.13}_{-0.11}$ are now established, whereas
the third angle is constrained by $\sin^2\theta_{13}<0.04$ and could
actually vanish.  ($2\sigma$ ranges, taken from
Ref.~\cite{Schwetz:2008er}.)  In the early universe, flavor
conversions are suppressed by large matter effects. Oscillations
driven by $\Delta m^2_{\rm atm}$ begin only at $T\sim10$~MeV whereas
those driven by $\Delta m^2_{\rm sol}$ begin as late as 3~MeV, still
early enough to achieve strong flavor conversions before
BBN~\cite{Lunardini:2000fy, Dolgov:2002ab, Wong:2002fa,
  Abazajian:2002qx}.  Therefore, all flavors must have similar
asymmetries at BBN and the restrictive limit on the $\nu_e$ degeneracy
parameter, conservatively $-0.04\alt\xi_{\nu_e}\alt 0.07$
\cite{Serpico:2005bc}, seemingly applies to all flavors, unless
oscillations are blocked, for instance by a hypothetical
neutrino-majoron coupling \cite{Dolgov:2004jw}.

Based on this reasoning it was often concluded that primordial
neutrino asymmetries could not cause a significant increase of the
cosmic neutrino density without violating the BBN bound on
$\xi_{\nu_e}$. Assuming that neutrinos indeed reach perfect kinetic
and chemical equilibrium, the possible increase of the cosmic
radiation density is small, $\Delta N_{\rm
eff}=\frac{30}{7}\,(\xi_{\nu}/\pi)^2+\frac{15}{7}\,(\xi_{\nu}/\pi)^4
\alt0.006$. Here, the usual ``effective number of neutrino
families'' measures the cosmic radiation density, after $e^+e^-$
annihilation, by $\rho_{\rm rad}/\rho_\gamma=1+(7/8)\,(4/11)^{4/3}\,
N_{\rm eff}\,$.

We stress, however, that Refs.~\cite{Dolgov:2002ab, Wong:2002fa,
Abazajian:2002qx} do not claim that neutrinos after flavor
conversions follow Fermi-Dirac distributions in terms of a common
degeneracy parameter and the temperature of the ambient plasma.
Refs.~\cite{Wong:2002fa, Abazajian:2002qx} showed that, if
$\theta_{13}=0$ and if one ignores collisions, synchronized MSW-like
flavor conversions driven by $\Delta m^2_{\rm sol}$ and
$\theta_{12}$ achieve comparable $\nu\bar\nu$ asymmetries among
all flavors, but this is not the same as kinetic and chemical
equilibrium. Ref.~\cite{Dolgov:2002ab} included collisions in a
schematic way, showing that the coherence among flavors was
efficiently destroyed. However, the transfer of entropy to the
electromagnetic plasma was not considered and the final neutrino
distributions not studied.

If all initial asymmetries have the same sign, oscillations alone are
enough to achieve approximate flavor equipartition and the BBN limits
on the $\nu_e\bar\nu_e$ asymmetry will apply to all flavors. A
dangerous case arises, however, if one begins with opposite
asymmetries in two flavors. Oscillations, with or without collisions,
will achieve approximate flavor equipartition separately among
neutrinos and antineutrinos, so in the end all asymmetries vanish.  On
the other hand, the original excess radiation remains.  The efficiency
of $\nu\bar\nu\to e^+e^-$ entropy transfer determines if the BBN
limits on the $\nu_e\bar\nu_e$ asymmetry still preclude a large
$\Delta N_{\rm eff}$. The main point of our paper is to study this
case that was not covered in previous treatments~\cite{Dolgov:2002ab,
  Wong:2002fa, Abazajian:2002qx}.

%%%%%%%%%%%%%%%%%%%%%%%%%%%%%%%%%%%%%%%%%%%%%%%%%%%%%%%%%%%%%%%%%%%%%%
{\em Numerical Approach.}---%
%%%%%%%%%%%%%%%%%%%%%%%%%%%%%%%%%%%%%%%%%%%%%%%%%%%%%%%%%%%%%%%%%%%%%%
We describe neutrino distributions by $3\times3$ matrices in flavor
space $\varrho_{\bf p}$ for each mode. The diagonal elements are the
usual occupation numbers whereas the off-diagonal ones encode phase
information. We need to solve the equations of motion
(EOMs)~\cite{Sigl:1993fn, McKellar:1994ja}
\begin{equation}
{\rm i}\,\dot\varrho_{\bf p}
=[{\sf\Omega}_{\bf p},\varrho_{\bf p}]+
C[\varrho_{\bf p},\bar\varrho_{\bf p}]
\end{equation}
and similar for the antineutrino matrices $\bar\varrho_{\bf p}$. The
first term on the r.h.s.\ describes oscillations with
\begin{equation}
{\sf\Omega}_{\bf p}=\frac{{\sf M}^2}{2p}+
\sqrt{2}\,G_{\rm F}\left(-\frac{8p}{3 m_{\rm w}^2}\,{\sf E}
+\varrho-\bar\varrho\right)\,,
\end{equation}
where $p=|{\bf p}|$, ${\sf M}$ is the neutrino mass matrix,
$\varrho=\int \varrho_{\bf p}\,{\rm d}^3{\bf p}/(2\pi)^3$ and similar
for $\bar\varrho$. The small baryon density implies that the dominant
matter term in the oscillation matrix is equal for neutrinos and
antineutrinos, where ${\sf E}$ is the $3\times3$ flavor matrix of
charged-lepton energy densities~\cite{Sigl:1993fn, Notzold:1987ik}.
The matrix $\bar{\sf\Omega}_{\bf p}$ for antineutrinos is the same
with ${\sf M}^2/2p\to -{\sf M}^2/2p$. The term proportional to
$\varrho-\bar\varrho$ is responsible for synchronized
oscillations~\cite{Dolgov:2002ab, Wong:2002fa, Abazajian:2002qx}.

Solving these EOMs with the full collision terms is numerically
challenging and not necessary for a first exploratory
study. Collisions have two main effects: They break the phase
coherence of mixed flavor states and they create or destroy neutrino
pairs. The first effect is efficiently implemented with $p$-dependent
damping factors for those $\varrho_{\bf p}$ elements that are
off-diagonal in the weak-interaction basis. Here we follow exactly the
earlier work of Dolgov et al.~\cite{Dolgov:2002ab} in that we use the
damping factors spelled out in Eqs.~(60) and~(61) of
Ref.~\cite{McKellar:1994ja}. Essentially one assumes that neutrinos
collapse into weak-interaction eigenstates each time they interact in
a process that distinguishes among flavors~\cite{Raffelt:1992uj}.

The crucial new ingredients are collisions and pair processes for the
diagonal $\varrho_{\bf p}$ elements, allowing the neutrino
distributions to achieve equilibrium with the ambient plasma.  With
small modifications we here use an existing numerical
code~\cite{Mangano:2005cc} where the collision integrals are solved
without approximation, using the same approach as
Ref.~\cite{Dolgov:1997mb}. For example, we obtain exact Fermi-Dirac
distributions if the relaxation time is large enough.

So in our implementation the diagonal elements change by collisions
and oscillations, whereas the off-diagonal elements change by
oscillations and damping. We have checked that the damping factors of
Ref.~\cite{McKellar:1994ja} and the collision rates used in the
code~\cite{Mangano:2005cc} are consistent with each other. Changing
the damping factors by as much as a factor of~2 has only a minimal
impact on the final $\Delta N_{\rm eff}$. Of course, our combined
treatment of oscillations, damping of flavor coherence, and collisions
and pair processes is only approximate and would be questionable if
one were to aim at a precision determination of $\Delta N_{\rm
  eff}$. However, our relatively simple extension of existing tools
provides a first reasonable estimate of the surviving $\Delta N_{\rm
  eff}$.

%%%%%%%%%%%%%%%%%%%%%%%%%%%%%%%%%%%%%%%%%%%%%%%%%%%%%%%%%%%%%%%%%%%%%%
{\em Opposite Equal Asymmetries.}---%
%%%%%%%%%%%%%%%%%%%%%%%%%%%%%%%%%%%%%%%%%%%%%%%%%%%%%%%%%%%%%%%%%%%%%%
Our main example is the ``dangerous case'' of two large but
opposite initial asymmetries. To be specific, we first consider an
exactly vanishing global lepton asymmetry. The initial asymmetry in
one flavor $\nu_\alpha$, defined in analogy to the baryon asymmetry,
is
\begin{equation}
\eta_{\nu_\alpha}=\frac{n_{\nu_\alpha}-n_{\bar\nu_\alpha}}{n_\gamma}
=\frac{1}{12\zeta(3)}\,
\left(\pi^2\xi_{\nu_\alpha}+\xi_{\nu_\alpha}^3\right)\,.
\end{equation}
A vanishing global asymmetry requires for the initial degeneracy
parameters $2\left(\pi^2\xi_{\nu_x}+\xi_{\nu_x}^3\right)+
\left(\pi^2\xi_{\nu_e}+\xi_{\nu_e}^3\right)=0$.

\begin{figure}[b]
\includegraphics[width=0.95\columnwidth]{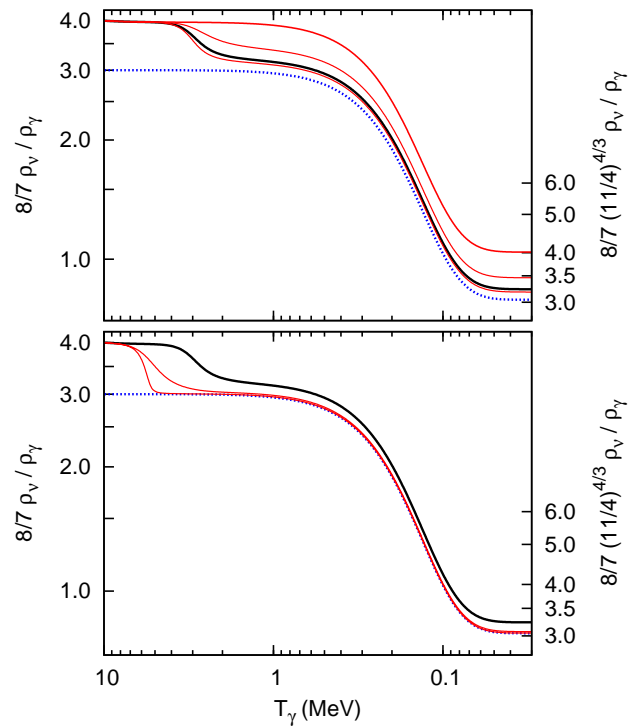}
\caption{\label{fig:neff}Evolution of the neutrino energy density.
  The vertical axis is marked with $N_{\rm eff}$, left before $e^+e^-$
  annihilation, right afterwards. Dotted lines: Baseline case without
  asymmetries. {\em Top panel}: $\theta_{13}=0$, the solid lines
  correspond to $\sin^2\theta_{12}=0$, 0.1, $0.3$, and 0.5 from top to
  bottom.  {\em Bottom panel:} $\sin^2\theta_{12}=0.3$, the top solid
  curve is $\sin^2\theta_{13}=0$, while the other solid curves
  correspond to $\sin^2\theta_{13}=0.04$ with normal (top) and
  inverted (bottom) hierarchy.}
\end{figure}

In Fig.~\ref{fig:neff} we show the evolution of
$\rho_\nu/\rho_\gamma$, properly normalized so that it corresponds to
$N_{\rm eff}$ at early and late times, first for the baseline case
without asymmetries (dotted lines).  The fast drop of
$\rho_\nu/\rho_\gamma$ at $T\sim0.2$~MeV represents photon heating by
$e^+e^-$ annihilation.  At late times the dotted lines end at $N_{\rm
  eff}=3.046$ instead of~3 because of residual neutrino
heating~\cite{Mangano:2005cc}.

Next we show (solid lines) the evolution for our main example where
$\xi_{\nu_x}=-0.64$ and $\xi_{\nu_e}=1.17$, corresponding to an
initial value $N_{\rm eff}=4.0$. The neutrino mass differences and
$\theta_{23}$ are fixed to their best-fit values.  In the upper panel
we assume $\theta_{13}=0$ and for the upper curve also $\theta_{12}=0$
(no oscillations), leading to an unchanged final $N_{\rm
  eff}=4.0$. The bottom curve is for $\sin^2\theta_{12}=0.5$ (maximal
mixing), showing that at $T_\gamma\sim 3$~MeV the neutrino density
depletes relative to photons, when oscillations driven by $\Delta
m^2_{\rm sol}$ begin and the flavour asymmetries are reduced.  The
main effect of collisions is to keep the neutrino spectra closer to
their equilibrium form: Physically the excess of entropy is
transferred from neutrinos to the electromagnetic plasma, cooling the
former and heating the latter.  Practically the same behavior is found
for the next curve representing $\sin^2\theta_{12}=0.3$, the
experimental best fit, where in the end $N_{\rm eff}=3.24$.  A further
curve is shown for the value $\sin^2\theta_{12}=0.1$, remaining far
from equilibrium until the end.  In the bottom panel we use the
measured $\sin^2\theta_{12}=0.3$ and repeat the case $\theta_{13}=0$
(top curve). The other two curves are for $\sin^2\theta_{13}=0.04$,
near the experimental limit, for the normal and inverted hierarchies.

As expected, a nonvanishing $\theta_{13}$ together with
$\Delta m^2_{\rm atm}$ drives the system towards equilibrium at a
time when collisions are still efficient. In our numerical example,
a significant deviation from equilibrium survives only for a very
small $\theta_{13}$. In this case the outcome depends on
$\theta_{12}$ as found in the original studies~\cite{Dolgov:2002ab,
  Wong:2002fa, Abazajian:2002qx}. A small $\theta_{12}$ does not lead
anywhere close to flavor equilibrium, a large value leads to
near--equilibrium.

\begin{figure}
\includegraphics[width=0.90\columnwidth]{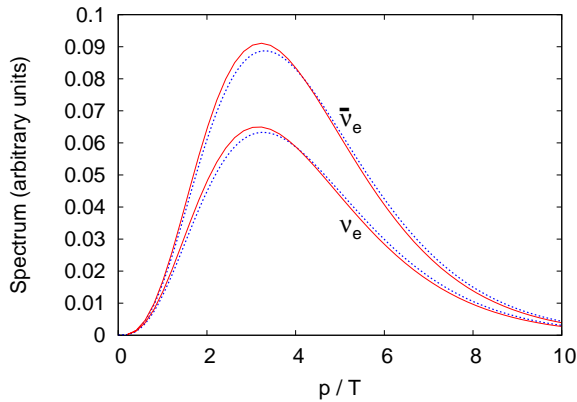}
\caption{\label{fig:spec}Final $\nu_e$ and $\bar\nu_e$ spectra for
  $\theta_{13}=0$ and $\sin^2\theta_{12}=0.3$. The
  dotted lines are Fermi-Dirac spectra producing the same
  $\nu_e\bar\nu_e$ asymmetry and the same energy density
  $\rho_{\nu_e}+ \rho_{\bar\nu_e}$.}
\end{figure}

For $\theta_{13}=0$ and $\sin^2\theta_{12}=0.3$ we show the final
$\nu_e$ and $\bar\nu_e$ energy spectra in Fig.~\ref{fig:spec}. We
compare them with Fermi-Dirac spectra that produce the same
$\nu_e\bar\nu_e$ asymmetry and the same energy density
$\rho_{\nu_e}+ \rho_{\bar\nu_e}$, implying $\xi_{\nu_e}=0.185$ and a
$T_{\nu_e}$ only 5\% higher relative to the equilibrium value.  These
spectra are much closer to kinetic equilibrium than
expected.

\begin{figure}[t]
%\vskip13cm
\includegraphics[width=0.9\columnwidth]{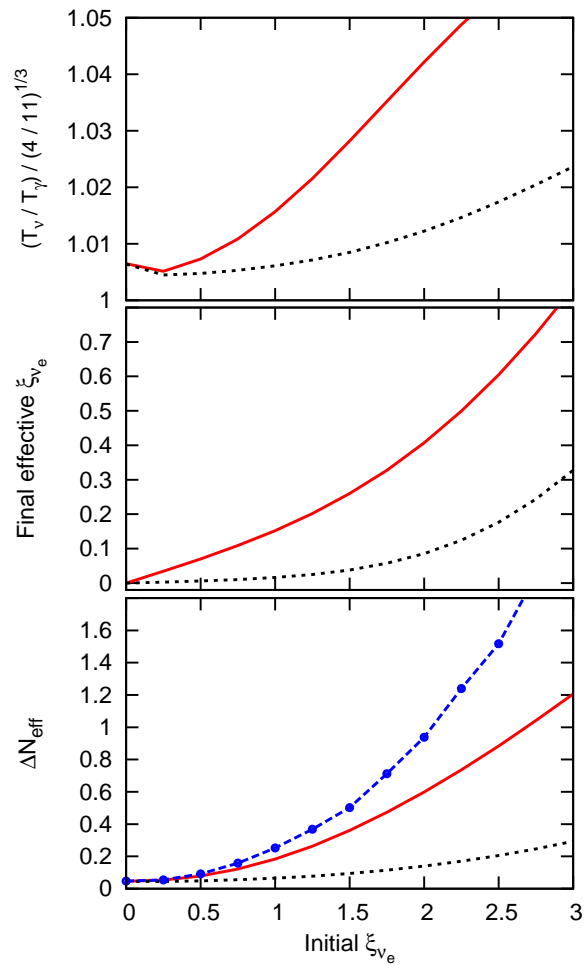}
\caption{\label{fig:scan}Parameters for the final $\nu_e$ and
  $\bar\nu_e$ spectra as well as the final $\Delta N_{\rm eff}$ as a
  function of the initial $\xi_{\nu_e}$. The solid lines correspond to
  $\theta_{13}=0$, the dotted lines to $\sin^2\theta_{13}=0.04$. In
  the bottom panel, the dashed line is the surviving $\Delta N_{\rm
    eff}$ when the final $\xi_{\nu_e}$ is in the range allowed by
  BBN.}
\end{figure}

%%%%%%%%%%%%%%%%%%%%%%%%%%%%%%%%%%%%%%%%%%%%%%%%%%%%%%%%%%%%%%%%%%%%%%
{\em Asymmetry Scan.}---%
%%%%%%%%%%%%%%%%%%%%%%%%%%%%%%%%%%%%%%%%%%%%%%%%%%%%%%%%%%%%%%%%%%%%%%
Next we study the final deviation from equilibrium for a range of
initial asymmetries $0\leq\xi_{\nu_e}\leq3$, assuming a
vanishing global asymmetry as before. In Fig.~\ref{fig:scan} we show
the final outcome in terms of the residual excess radiation density
$\Delta N_{\rm eff}$ and $\xi_{\nu_e}$ and $T_{\nu_e}$ chosen for a
Fermi-Dirac distribution that approximates the final $\nu_e$ and
$\bar\nu_e$ spectra such that we reproduce the numerical asymmetry and
energy density. We show these results for the experimental best-fit
value $\sin^2\theta_{12}=0.3$ and for $\theta_{13}=0$ as well as
$\sin^2\theta_{13}=0.04$.

%\begin{figure}[ht]

Qualitatively we confirm that kinetic equilibrium is rather good in
that the final neutrino temperature is close to what it should be in
equilibrium. Moreover, we also find that the value of $\theta_{13}$ is
crucial in determining how close the system comes to equilibrium,
which however is never perfect for any value of $\theta_{13}$. If
$\sin^2\theta_{13}\alt 10^{-3}$, the outcome is very close to a
vanishing  $\theta_{13}$.

We note that for large initial $\xi_{\nu_e}$ the final electron
neutrino degeneracy can be so large that it is excluded by
BBN. However, we are not bound to the assumption of an initially
vanishing global asymmetry, but we can use slightly more negative
initial $\xi_{\nu_x}$ values such that the final $\xi_{\nu_e}$ is
pulled into the range allowed by BBN, yet a significant excess
radiation density survives as shown in in Fig.~\ref{fig:scan}, being
quite similar to $\Delta N_{\rm eff}$ for the simple case of a
vanishing global asymmetry. In other words, up to small adjustments of
initial conditions, the simple case of a vanishing global asymmetry
gives us a fair idea of a possible surviving residual radiation
density.

%%%%%%%%%%%%%%%%%%%%%%%%%%%%%%%%%%%%%%%%%%%%%%%%%%%%%%%%%%%%%%%%%%%%%%
{\em No Initial $\nu_e\bar\nu_e$ Asymmetry.}---%
%%%%%%%%%%%%%%%%%%%%%%%%%%%%%%%%%%%%%%%%%%%%%%%%%%%%%%%%%%%%%%%%%%%%%%
Another initial configuration for a vanishing global asymmetry is
$\xi_{\nu_e}=0$ and $\xi_{\nu_\mu}=-\xi_{\nu_\tau}$. If the latter are
exactly equal but opposite, $\nu_\mu\nu_\tau$ oscillations driven
by $\Delta m^2_{\rm atm}$ and $\theta_{23}$ are suppressed by
neutrino-neutrino interactions in that the synchronized oscillation
frequency vanishes exactly~\cite{Dolgov:2002ab}. However, oscillations
driven by $\Delta m^2_{\rm sol}$ and $\theta_{12}$ achieve approximate
equilibrium, the transformations beginning at $T\sim3$~MeV as
before. We have not studied this case in detail because it is
numerically more challenging. We note that ``blocking'' the
$\nu_\mu\nu_\tau$ oscillations requires a very precise cancellation
and thus a more special case than before where an exactly vanishing
global asymmetry was only a choice of convenience.

%%%%%%%%%%%%%%%%%%%%%%%%%%%%%%%%%%%%%%%%%%%%%%%%%%%%%%%%%%%%%%%%%%%%%%
{\em Conclusions.}---%
%%%%%%%%%%%%%%%%%%%%%%%%%%%%%%%%%%%%%%%%%%%%%%%%%%%%%%%%%%%%%%%%%%%%%%
The first studies of primordial neutrino flavor equilibration were
concerned with the role of the ``solar'' neutrino mixing parameters
and found that approximate global flavor equilibrium is
only achieved for the now-established ``large mixing angle
solution'' \cite{Dolgov:2002ab, Wong:2002fa, Abazajian:2002qx}.
While this conclusion is correct, we here stress that it must not be
over interpreted.  Oscillations driven by the solar parameters take
place so late ($T\sim 3$~MeV) that neither perfect flavor
equilibrium nor thermal equilibrium with the ambient plasma is
achieved, although a large value for $\theta_{13}$ helps to come
closer because flavor transformations begin earlier. (If
$\sin^2\theta_{13}\alt10^{-3}$ it plays no significant role.)
For $\theta_{13}=0$ we have constructed an explicit
example with nearly equal but opposite asymmetries between $\nu_e$
and the other flavors such that the BBN limit on $\xi_{\nu_e}$ is
satisfied, yet a large $\Delta N_{\rm eff}$ survives.

Precision cosmology has advanced in such strides that today one can
derive nontrivial bounds on the cosmic radiation density that however
still allow for $\Delta N_{\rm eff}$ of a few~\cite{Mangano:2006ur,
  Hamann:2007pi, Bernardis:2007bu, Komatsu:2008hk, Popa:2008tb,
  Ichikawa:2008pz}. It is conceivable that in future one can do much
better and constrain or detect a value of $\Delta N_{\rm eff}$
significantly smaller than unity~\cite{Hamann:2007sb,
  Bashinsky:2003tk, Hannestad:2006as, Perotto:2006rj,
  Friedland:2007vv}, for instance with very precise CMB data from
Planck.  Such results remain complementary to BBN limits on
$\xi_{\nu_e}$. If precision cosmology were to turn up a nonvanishing
$\Delta N_{\rm eff}$, it could be a remnant of a primordial neutrino
asymmetry, although the exact interpretation would strongly depend on
$\theta_{13}$. In this sense, future measurements of $\theta_{13}$ are
important for the interpretation of cosmological precision data.

In summary, the established neutrino mixing parameters together with
BBN constraints on $\xi_{\nu_e}$ typically assure that the surviving
excess radiation density is immeasurably small, in agreement with the
previous literature. However, imperfect kinetic and chemical
equilibrium achieved by the solar mixing parameters allows for
exceptions. In particular, large but opposite primordial asymmetries
of two flavors can be fine--tuned such that the $\xi_{\nu_e}$ limit is
respected, yet a large excess radiation density survives with $\Delta
N_{\rm eff}$ of order unity or larger that may be detectable in future
cosmological precision data.

%%%%%%%%%%%%%%%%%%%%%%%%%%%%%%%%%%%%%%%%%%%%%%%%%%%%%%%%%%%%%%%%%%%%%%
% Acknowledgements %%%%%%%%%%%%%%%%%%%%%%%%%%%%%%%%%%%%%%%%%%%%%%%%%%%
%%%%%%%%%%%%%%%%%%%%%%%%%%%%%%%%%%%%%%%%%%%%%%%%%%%%%%%%%%%%%%%%%%%%%%

{\em Acknowledgments.}---We thank Y.Y.Y.~Wong for comments on the
manuscript. This work was partly supported by the DFG (Germany) grant
TR-27 ``Neutrinos and Beyond,'' the Cluster of Excellence ``Origin and
Structure of the Universe,'' the European Union (RT Network
MRTN-CT-2004-503369), and the Spanish grants
FPA2005-01269/FPA2008-00319 (MICINN) and BEST/2008/164 (Generalitat
Valenciana). TP was supported by a Spanish FPU grant. SP thanks the
MPI for Physics in Munich for hospitality and financial support for
his visit when this work was completed.

%%%%%%%%%%%%%%%%%%%%%%%%%%%%%%%%%%%%%%%%%%%%%%%%%%%%%%%%%%%%%%%%%%%%%%
%%%  Bibliography  %%%%%%%%%%%%%%%%%%%%%%%%%%%%%%%%%%%%%%%%%%%%%%%%%%%
%%%%%%%%%%%%%%%%%%%%%%%%%%%%%%%%%%%%%%%%%%%%%%%%%%%%%%%%%%%%%%%%%%%%%%

%%%%%%%%%%%%%%%%%%%%%%%%%%%%%%%%%%%%%%%%%%%%%%%%%%%%%%%%%%%%%%%%%%%%%%
\end{document}